# RadioNet3 Study Group White Paper on: The Future Organisation and Coordination of Radio Astronomy in Europe.

QueSERA (WP2, TASK1)

M.A. Garrett (chair), P. Charlot, S.T. Garrington, H-R Klöckner, H.van Langevelde, F. Mantovani, A. Russel, K. Schuster, R.C. Vermeulen, & A. Zensus (QSG Study Group).

December 2015

**Summary**

**The QueSERA Study Group (QSG) have been tasked by the RadioNet Board to produce a White Paper on the future organization and coordination of radio astronomy in Europe. This White Paper describes the options discussed by the QSG, and our conclusions on how to move forward. We propose, that as a first step, "RadioNet-work" be established as an entity that persists between EC contracts, and that takes responsibility for preparing or coordinating responses to EC opportunities specific to the field of radio astronomy research infrastructures. RadioNet-work should provide a safety net that ensures that cooperation and collaboration between the various radio astronomy partners in Europe is maintained with or without EC funding.**

*Introduction*

The field of Radio Astronomy is a vibrant one, and changing rapidly. Many existing telescopes are undergoing significant upgrades, and new, large-scale facilities such as ALMA and the SKA are set to have a major impact in the field, especially so in Europe. European collaboration in radio astronomy is stronger than ever, with RadioNet being perhaps the most prominent and visible example.

This QSG White Paper focuses on the topic of the future coordination of radio astronomy in Europe, presenting a range of options and a motivated recommendation on how to move forward. The paper also takes into account commentary from other relevant stakeholders. It represents a synthesis of discussions and input that have taken place over the last 3 years (a list of key meetings is provided in appendix 1).

*European radio astronomy landscape*

As a pre-requisite, the QSG has reviewed the current status of Radio Astronomy in Europe, and how this is likely to develop over the coming decade. The field is currently dominated by national facilities and there is a strong ambition and mandate for these to continue as front-rank instruments for at least the next decade and in many cases beyond this time frame.

The national facilities often combine together to form internationally distributed (interferometer) networks e.g. the EVN, Global VLBI, mm-VLBI, EPTA etc. Indeed several international organisations are involved in operating European radio astronomy facilities (ESO, IRAM, JIVE and the ILT). These entities have played a crucial role in making top-class radio astronomy facilities openly available to the full European and international communities.

Europe is blessed with an array of standalone, single-dish radio telescopes operating at cm, mm and sub-mm wavelengths. These are generating a wide range of science, with particular highlights being studies of pulsars (including pulsar timing), gravity tests and the chemistry of molecular clouds and dust in both regions of our own galaxy and extragalactic systems. Continuous maintenance and investment in these instruments have maintained them at the forefront of the field. In particular, new developments such as the upcoming installation of the NIKA2 instrument on the IRAM 30-metre and the development of both wide-band receivers and cryogenically cooled Phased Array Feeds (PAFs), ably demonstrate how relatively modest upgrades in existing single-dish telescopes can deliver high-impact, state-of-the-art scientific results. Interferometers such as e-MERLIN, WSRT, PdBI are state-of-the-art instruments with broad international communities and on-going upgrade programmes. e-MERLIN seeks seamless integration with the EVN, enabling the study of individual radio sources on many different angular scales. The upgrade of the WSRT to Phase Array Feeds is now underway, and it is expected to provide a legacy survey of the Northern sky in both the radio continuum and neutral hydrogen.

Significant growth is foreseen for VLBI in general - ALMA for example should have a phased-up VLBI capability soon, and SKA1-mid is expected to also participate as a very sensitive element of a global array. The restriction of SKA1-mid to baseline lengths of a few hundred km at best, suggests the demand for European and global VLBI will only increase over the coming decades. The suite of cm/mm facilities in Europe has recently received a boost with the commissioning of the SRT in Sardinia. The recent establishment of JIVE as an ERIC (European Research Infrastructure Consortium - a legal entity initiated and now recognised by the EC and its member states) is another important step forward for VLBI in Europe.

NOEMA, the successor to the Plateau du Bure observatory, will be the most powerful mm radio telescope in the Northern Hemisphere - a good demonstration of the scientific need to retain an "all-sky" capability with ALMA located in the Southern hemisphere. NOEMA is a platform on which new advanced technologies can be easily implemented, and in which a complementary advantage can be delivered to the European community as they compete for time on ALMA with their US and Japanese colleagues. The QSG believes that NOEMA represents a good example of how cm-wavelength facilities in Europe might also develop in the era of the SKA.

At the other end of the radio spectral domain, the International LOFAR Telescope (ILT) is seen as a major new European initiative that continues to expand, currently extending from the Netherlands into France, Germany, Sweden and the UK. A sixth German LOFAR station has recently been completed near Hamburg, and 3 new stations are now under construction in Poland. Ireland is also expected to build a LOFAR station at Birr Castle. As in the case of the EVN, LOFAR's long-baseline capability will also greatly exceed that of SKA1-low, and the long-term future of this infrastructure is also secure. In addition, LOFAR's capabilities as a major instrument for cosmic ray research continues to grow – this also links the radio astronomy community to the high-energy astro-particle physics community – a link that is likely to become more important in the future via projects such as ASTERICS. Significant upgrades to LOFAR can also be expected in the coming decade, including a possible re-use of the infrastructure for higher frequency Aperture Array facilities.

The SKA is making good progress and is now well into the design phase of SKA-1. It is clear that the majority of the international SKA partners wish to see the project governed via an Inter-Governmental Organisation. The way in which Europe is represented in the project is currently a topic of some discussion – while some European countries wish to be represented within the SKA IGO as individual national partners, others would like to see Europe itself represented via an appropriate organisation e.g. ESO. A role for ESO in the SKA project is currently being explored but no conclusions are yet to be drawn. The number of European countries aspiring to join the SKA will therefore impact on a possible role of ESO. The role of JIVE in the SKA is also still to be fully determined at this stage but at least the VLBI component of SKA1-mid, and the link this makes with both the fledgling African VLBI Network (AVN) and the EVN is likely to require its active support. VLBI can be an important vehicle for consolidating the strategic link between the radio astronomy community in (South) Africa and Europe.

In addition, several radio astronomy institutes in Europe are interested in participating and supporting a European SKA Regional Centre (SRC). Such a centre is a key aspect of the operational model now being developed by the SKA Office. The exact structure of such an entity is still to be fully explored but it probably contains both centralised and distributed elements. The mission of an SRC might include one or more aspects of data management, data access, scientific support, science exploitation and technical R&D for SKA upgrades (including SKA Phase 2). An organizational structure similar to the ALMA Regional Centres (ARCs) and associated arclets is one possible model to follow (at least in terms of astronomical user support). Creating such an organisation is important in order for European scientists to fully exploit the SKA, and can help to retain and grow the scientific and technical radio astronomy communities here in Europe. A European Working Group has recently been established to consider how a federated approach to SKA Science Data Processing might be realised. The SKA Office is also working hard to establish its own vision of an SKA Observatory operational model that includes external contributions/participation.

*Forward look*

From the QSG discussions, it is clear that there is a strong ambition to maintain a major European radio telescope capability in the Northern hemisphere, in addition to strong involvement in both ALMA and the SKA in the South. The QSG concluded that:

(i) it is essential to nurture and indeed further grow the vibrant European Radio Astronomy science community in the era of both ALMA & SKA,

(ii) it is critical that the impressive technical and engineering expertise in European radio astronomy is retained,

(iii) efforts to protect passive use of the radio spectrum for European and global radio astronomy facilities must be vigorously maintained,

(iv) there is broad interest in hosting a SKA Regional Centre in Europe covering a wide-range of distributed support activities and interests – this development is seen as crucial and complementary to the 2 previous statements,

(v) national institutes have strong ambitions to maintain and indeed upgrade current (complementary) facilities on time scales that extend well beyond the current decade,

(vi) the NOEMA upgrade of PdB is considered to be an excellent example of how European m/cm wavelength radio astronomy might also further develop in the SKA era,

(vii) VLBI both at mm and cm wavelengths represents a strong scientific case when augmented with ALMA and the SKA as additional network elements.

*(viii)* if ESO's role in European radio astronomy should grow, the European radio astronomy community must be engaged in this process.

### *Stakeholder input*

Two sources of important input to the development of this White Paper included the ERTRC (European Radio Telescope Review Committee) report and feedback from the mid-term review of the RadioNet3 project. In this section, we highlight specific recommendations that we believe this White Paper addresses in its conclusions.

### *ERTRC report*

Several years ago, ASTRONET took the initiative to set up an independent European Radio Telescope Review Committee (ERTRC). The ERTRC draft report appeared in July 2013 and the community (including RadioNet) provided input to the document and the associated list of recommendations. In 2014, ASTRONET organised a Round Table meeting between ASTRONET, RadioNet3 and Go-SKA, in response to the ERTRC draft report. The final ERTRC report was published in June 2015 with significant input provided by RadioNet and its partners.

In general, RadioNet endorses the ERTRC report. The conclusions of the report are complementary to the vision for radio astronomy in Europe also presented in this white paper, and there is significant overlap between our own conclusions and the recommendations of the ERTRC.

Of particular relevance to the activities of the QSG were the following ERTRC recommendations:

*ERTRC report, section 12.2, recommendation 7:*

*"We recommend that local and national radio institutes remain independent, as local support and expertise centres for radio astronomy, but that their joint activities, such as EVN and RadioNet, become more robustly and permanently organised and funded (but not through the same body that organises the European participation in the SKA). (16, 18)"*

*ERTRC report, section 12.3, recommendation 16b*

*"We are concerned that current arrangements for European collaboration in radio astronomy are not robust and secure enough to safeguard such collaboration towards the future, especially given that this collaboration will need to become more intense. (Ch.10) a. …, b. The continuity of activities such as RadioNet, YERAC, and CRAF, needs to be guaranteed".*

*Mid-Term Review of RadioNet*

The mid-term review of RadioNet has also provided some input to this White Paper. The advice to tread carefully was well made. The suggestion of a "light" organisation for European Radio Astronomy is one that this paper has taken on-board directly.

**Options for European Radio Astronomy Coordination**

Several possible options were studied by the QSG but two were given particular attention:

1) A gentle evolution of the current RadioNet collaboration with a mandate for to establish a new entity "RadioNet-work" charged to coordinate European Radio Astronomy in some specific, pre-agreed areas of common interest, all based upon a "light" MoU,

2) The establishment of a new (or the adoption of an existing) legal entity/organisation for the central coordination of European Radio Astronomy.

*Conclusions*

Option 1: The need to extend the current collaboration in European radio astronomy is a strong one. RadioNet has developed a broad range of common activities that need to be maintained with or without EC funding – perhaps the best example is the need to lobby for and prepare a response to, EC calls that specifically target the field of radio astronomy. Other essential activities include: YERAC, CRAF, Outreach (on a European scale), a representative function for European RA, S&T coordination in RA etc.

A natural first step (option 1), is to establish "RadioNet-work" as a body that is persistent (i.e. it exists at all times, irrespective of whether it is in receipt of EC funding), and it is active in maintaining a minimum level of key activities (see above) between (or in the absence of) EC funding cycles. A low-barrier is desirable with respect to participation in RadioNet-work but a modest annual contribution from the various partners may be foreseen. The location of the RadioNet-work office is expected to be mobile and perhaps even distributed but it would logically follow that the major part would reside at the RadioNet coordinators' place of work.

Option 2: the creation of a new (or existing) legal entity to coordinate European radio astronomy was seen as premature at this stage. The discussion within the QSG itself, suggests that it would be very difficult for a consensus to form around such an initiative across all the RadioNet partners. It was also recognised that establishing a legal entity would demand a level of funding and resources that is not readily available, even at funding agency level.

For both models, the aspect of stronger centralised control raised some worries, especially w.r.t. maintaining a broad and geographically distributed radio astronomy expertise and presence across Europe. In testing this aspect against the various options, it was concluded that option 1 (with a much lighter approach and limited ambition) presented the minimum risk.

In short, this white paper recommends that the RadioNet board endorses the establishment of RadioNet-work. The creation of such an entity must be fully in line with, and supportive of, the ambition to submit a new RadioNet4 proposal for the upcoming EC opportunity (INFRAIA-01-2016/2017: Integrating Activities for Advanced Communities). Appendix 3 presents a draft of a possible mandate for the RadioNet-work initiative.

**Appendix 1**

ToR of the QSG (see RadioNet-3 deliverable D2.2). An excerpt of D2.2. is provided below.

**RadioNet3 Study Group: Organisation of Radio Astronomy in Europe.**

In its interaction with policy makers, the obvious issue for European radio astronomy is the future of its structure as a whole. RadioNet3 as a project and a consortium is rather loosely organized and has no coordinated long-term perspective on the European scale. Especially with the advent of the SKA, a natural question to ask is whether a new legal entity for radio astronomy is required within Europe. The potential role of existing vehicles (e.g. the current ESKAC collaboration, ESO, or a future JIVE-ERIC) is also relevant here. The deliverables and milestones of this work package take the form of face-to-face meetings including invitations to relevant external parties as appropriate, plus a final position paper.

This will produce a roadmap for existing RadioNet3 facilities that

- Recognises the impact ALMA and the SKA will have in the field, and builds and responds to the current ASTRONET review process,
- Defines the future role of existing facilities in the Northern hemisphere (incl. VLBI),
- Identifies an appropriate model for SKA scientific (user) support, that incorporates lessons learned from the ALMA experience,
- Establishes a clear vision on how the European radio astronomy community should formally organise itself in the coming decade,
- Addresses the need for future European scale integrating activities beyond RadioNet3 and consider how these should be funded.

**Background**

1. The field of Radio Astronomy is changing rapidly. In Europe, major new telescopes such as ALMA, LOFAR and the SRT are in the commissioning phase, existing telescopes are benefiting from substantial upgrades (e.g. e-VLBI, *e*-MERLIN, PdBI- NOEMA etc.) and the SKA has entered the pre-construction phase. The field is widely considered to be flourishing around the world, and new regions are emerging to play an important role in its future course (e.g. China and South Africa).

2. Over the last 3 decades, different structures have been set up to organise and coordinate a wide range of key but specific activities in radio astronomy at a European level - these include the EVN & JIVE (VLBI), the ILT (LOFAR), ESKAC, ESO, IRAM, RadioNet, NEXPReS, GO-SKA (EC Framework Programmes) and most recently the SKA Organisation (SKA). In general, these structures have worked well and delivered in terms of organising the community and providing access to state of the art astronomical facilities.

3. The plurality of this multi-faceted approach has often provided added value to the community. The existence of many different radio astronomy organisations in Europe has led to a broad spread of both scientific and technical expertise. However, the inter-relationships between the various radio astronomy entities in Europe and their dependence on each other is complex. The lack of a tighter coordination of Radio Astronomy as a whole may already limit what can be achieved on the European scale, and future funding may require a more centralised approach. As radio astronomy activities are being reviewed externally by others, it makes sense for the INFRA-2011-1.1.21 *RadioNet3* RadioNet community itself to also consider whether we can better organise ourselves in the future or whether the current diverse system is already the optimal approach.

4. A proper analysis of these questions requires a full understanding of the likely development of radio astronomy on a European and national scale. RadioNet has contributed to the ASTRONET road mapping process, and a focused review of long wavelength radio astronomy facilities conducted by ASTRONET is underway. Rather than to repeat these exercises, the Study Group will make as much use as possible of the current ASTRONET Radio Telescope review. However, its important that the RadioNet community has its own view on an agreed roadmap for Radio Astronomy in Europe. Understanding the future development of Radio Astronomy at both the national and European level is a pre-requisite before appropriate collaboration models can be fully assessed. This can be achieved also as part of our community's contribution to the ASTRONET roadmap update currently underway.

**Scope**

The main objective is to formulate a clear vision for the organisation and coordination of European Radio Astronomy - a vision that should ensure sustainable growth in the field for at least the next 10 years and that can be agreed by the full radio (m/cm/mm) community.

Two key issues will be addressed:

- Understanding the roadmap for Radio Astronomy over the coming decade on both national and European scales,

- The future coordination of Radio Astronomy in Europe taking into account a diversity of approaches that may need to be applied across the full field.

**Outputs**

A Study Group will be established to produce a final position paper on the future coordination of Radio Astronomy in Europe. The Group will consider as input the Roadmap for existing RadioNet3 facilities with a particular focus on those that require collaboration on a European scale. During the lifespan of the Group, and to facilitate its deliberations, a number of working papers and policy briefs may be prepared for discussion at both Group and RadioNet3 Board level.

**Organisation of Work**

*Study Group*

Membership of the Study Group will include: the RadioNet3 coordinator, the QueSERA Work Package leader and representatives of TNA facilities that are distributed or multi-national in scope (e.g. ESO, EVN, IRAM and ILT). Meetings of the Group will be chaired by the task leader and open to all RadioNet3 Board members. External Guests and other expert stakeholders (e.g. ESO, SKA, ASTRONET etc.) may also be invited to meetings at the invitation of the chair. The Group reports to the RadioNet3 Board. The RadioNet3 Board will approve by consensus any outcomes of the Group, in particular the position paper. The RadioNet3 Board will be required to approve the public dissemination of any outputs from the task.

*The Secretariat*

A secretariat will be provided by RadioNet- to support the Group in terms of meeting logistics etc. and to take minutes of the meetings. The Group will consult widely with funding agencies, the EC, existing governance bodies in Radio Astronomy, the Radio Astronomy community, individuals, organisations and any other relevant entities with a view to collecting information and generating ideas relevant to the work of the Group.

**Timeline**

The Group will complete its work by the approval by the Board of a final position paper in June 2015. Three face-to-face meetings of the Group will be organised during this period. Additional meetings may be organised by electronic means with varied participation, as appropriate. The Group may establish further working groups led by one or two members to prepare input on specific topics for its consideration. The Study Group will explore all possible opportunities to promote the outcomes of the work and to engage with the radio astronomy community.

**Appendix 2**

*QSG meetings*

Face-to-face QSG meetings were open to all RadioNet3 board members, in addition to members of the QSG itself.

Notes from the meeting have been made by the chair, and are available via the various RadioNet3 deliverables.

The chair recorded summaries of the major discussions and noted decisions/action items. These have formed part of the deliverable to the EC.

A list of f2f meetings, telecons and other relevant events include:

14 October 2013, MGP Office (Rue Royale 225-227; B-1210 Bruxelles).
31 January 2014, MGP Office (Rue Royale 225-227; B-1210 Bruxelles).
24 October 2014, MGP Office (Rue Royale 225-227; B-1210 Bruxelles).
15 January 2015 (telecon)
24 February 2015, Bordeaux (face-to-face meeting + MAG via telecon)
25 February 2015, Bordeaux – presentation of draft white paper finding to RadioNet Board and EC Project Officer.

**Appendix 3**

In this appendix we present the latest (5.10.2015) RadioNet-work concept as drafted by the RadioNet coordinator Anton Zensus et al.

Since the draft is likely to evolve over the coming months, we also present a password protected link, available to the Board at:

http://www.radionet-eu.org/radionet3wiki/doku.php?id=na:management:radionet-work_scenario_-_preparation

# RadioNet-*work*

*(version 5.10.2015)*

ASTRONET/ERTRC Report – Recommendation 7:

*[…] We recommend that local and national radio institutes remain independent, as local support and expertise centres for radio astronomy, but that their joint activities, such as EVN and RadioNet, become more robustly and permanently organised and funded (but not through the same body that organises the European participation in the SKA).*

**Background**

RadioNet is an established brand in Europe, coordinating substantial trans-national access, and comprehensive networking and joint research activities, with EU funding of currently 9.5 Mio Euros over 4 years (2012-2015), and individual national contributions. RadioNet also is recognized as a de facto representative of common interests of the European radio astronomy community of facility operators, telescope users and researchers, and engineers. A new proposal is being prepared (as of 10/2015) for the funding period 2017-2020.

The European Commission expects from the advanced communities to develop a plan for long- term sustainability. It is anticipated that a strategic roadmap for the future research infrastructure developments as well as a sustainability plan beyond the EC grant lifecycle will be prepared. Hence RadioNet should ultimately become independent from the EC funding, and become able to organize the required institutional funds to sustain community efforts out of national funding streams.

While RadioNet has been successful in making a case for coordination of RadioNet in Europe, one must recognize that the driving motivation for coming together has been the expectation and benefit of European funding supplementing or in some cases replacing reduced national funding for telescope operations and for technical development. The absence of such a funding opportunity would likely after a while lead to loss of purpose and lack of motivation to keep up with the inevitable overhead of maintaining sometimes quite onerous coordinated activities. These might then well be replaced by subgroups losing interest and pursuing other suitable opportunities. Furthermore, at present it appears unlikely that a new central European entity/organization will emerge that would take over most of the roles RadioNet is serving.

Past and present experiences (EVN, JIV ERIC, ESKAC) highlight that the multitude of funding sources and regulations in different countries present substantial difficulties in creating and maintaining even a modest joint fund with regular contributions of all partners. Thus an organization based on modest contributions through a Memorandum of Understanding (MoU) seems to be the most suitable solution. This is referred to here as RadioNet-*work*, anticipating that ultimately this will take the name RadioNet.

**DRAFT Principles to be agreed for the Creation of RadioNet-work DRAFT**
*(once agreed, these principles will be translated into a MoU)*

- [Preamble] The signing parties agree on their joint interest
    - To foster collaborative activities among radio astronomy institutes and laboratories in Europe, in support of their common interests.
    - To lobby for and coordinate joint applications to EU funding opportunities in radio astronomy
    - To maintain a modest, self-funded collaborative programme of networking activities (e.g., to fund YERAC, schools, science and engineering meetings, and travel for selected CRAF and Time Allocation Committee meetings).
    - To coordinate a joint representation in the European research are (e.g., ASTRONET, SKA)

- [Therefore…] The signing Parties agree
    - To form by Memorandum-of-Understanding a *RadioNet-work Consortium,* with the objective to implement the above goals.
    - RadioNet will have full members, committed to the joint programme and contributing financially to a modest annual budget. The members have full voting rights.
    - There will also be a class of Associate Membership, which will allow participation in the activities of RadioNet at a less formal level and without a fee.
    - The members form a Board populated with one representative and vote per member. The Board is the ultimate decision making body and will agree on appropriate terms- of-procedure.
    - Associate members will be invited to send a representative to Board meetings, without voting rights.
    - The Board will meet in person at least once a year, with additional meetings possible in person or by teleconferencing
    - The Board will elect a Chair and vice-Chair for terms of 3 years, and may

appoint a secretary.

- The Board will appoint a Coordinator, normally the same person as the Coordinator of the corresponding EU programme.
- An "Office" will be established to coordinate activities, staffed at least of the Coordinator and an Assistant. Ideally this would be in an independent location (e.g., Brussels), with some key personnel; in reality, limited funds will require that this may need to be a modest in-kind or pro-bono contribution from the Coordinator's organisation.
- RadioNet-work will maintain a bank account. A member organisation establishes this as an audited account on behalf of the Consortium.
- The initial Annual Contribution will be 10,000 Euros per member.
- Changes to the initial annual contribution will be approved by the Board unanimously and depend on the Board approved budget and the number of payers, and in any case approved by the Board.
- The initial duration of the agreement is for 5 years.
- The operation of RadioNet-work will begin on 1 January 2016.
-

• **DRAFT Initial Budget**

The critical effectiveness of RadioNet-work has been identified as basic management office, consortium meetings and representation, maintenance of the networking events (YERAC, schools, engineering meetings). Therefore the rudimentary annual budget of the RadioNet-work can be estimate as follows:

1. Cost of the office (50.000€):

    - Chair (10% FTE in-kind contribution);

    - Assistant (0,5 FTE)

    - Office rent (in case this is not allocated at the partner's institute)

2. Cost of the Consortium meeting (10.000€)

    - organisation cost

    - travel of the officers and and invited persons

    - lobbying activities

3. Cost of the networking meetings (40.000€):

    - YERAC,

    - ERIS/ single dish schools

    - Engineering meetings (TOG/GMVA, CRAF, etc.)

- Selected TAC meetings, as mandated for EU-funded transnational access